\documentstyle[12pt]{article}
\input{psfig}
\newcommand{\pom}{I\!\! P}
\begin{document}
\vspace*{-6ex}
{\hfill
\fbox{Rockefeller University Report: RU 97/E-26}}
\vglue 1.5cm
\begin{center}

{\large\bf
Comments on the Erhan-Schlein model of damping the pomeron flux 
at small x-pomeron}

\vglue 0.3cm
{K.  GOULIANOS\\ }
(dino@physics.rockefeller.edu)
\vglue 0.2 cm
\baselineskip=13pt
{\em The Rockefeller University, 1230 York Avenue\\
 New York, NY 10021, USA\\}
\vskip0.25in

\today

\end{center}

\centerline{ABSTRACT}

{\rightskip=10pc
\leftskip=10pc
We explore the theoretical and experimental consequences of a model 
proposed by Samim Erhan and Peter 
Schlein for unitarizing the diffractive amplitude by 
damping the pomeron flux at small x-pomeron and conclude that the model 
is unphysical and contradicts well established experimental data.
}
\section{Introduction}
In Regge theory, the high energy behaviour of hadronic cross sections is 
dominated by pomeron exchange~\cite{KG,DLT}. 
For a simple pomeron pole, the $pp$ elastic, 
total and single diffraction dissociation cross sections are given by
\begin{equation}
\frac{d\sigma_{el}}{dt}=\frac{\beta^4_{\pom p}(t)}{16\pi}\;
{\left(\frac{s}{s_0}\right)}^{2[\alpha(t)-1]}
=\frac{\beta^4_{\pom p}(t)}{16\pi}\;
{\left(\frac{s}{s_0}\right)}^{2\alpha' t}\;{\left(\frac{s}{s_0}\right)}^
{2\epsilon}
\label{elastic}
\end{equation}
\begin{equation}
\sigma_T=\beta^2_{\pom p}(0)
\left(\frac{s}{s_0}\right)^{\alpha(0)-1}=\beta^2_{\pom p}(0)
{\left(\frac{s}{s_0}\right)}^{\epsilon}
\label{total}
\end{equation}
\begin{equation}
\frac{d^2\sigma_{sd}}{d\xi dt}=
\frac{{\beta_{\pom p}^2(t)}}{16\pi}\;\xi^{1-2\alpha(t)}
\cdot \left[ \beta_{\pom p}(0)\,g(t)
\;\left(\frac{s'}{s_0}\right)^{\alpha(0)-1}\right]
\equiv f_{\pom /p}(\xi,t)\cdot
\left[\sigma_T^{\pom p}(s',t)\right]
\label{diffractive}
\end{equation}
where $\alpha(t)=1+\epsilon+\alpha' t$ is the pomeron Regge trajectory, 
$\beta_{\pom p}(t)$ is the  coupling of the pomeron to the proton,
$g(t)$ the triple-pomeron coupling,
$s'$ the $\pom-p$ center of mass energy squared,
$\xi\equiv x_{\pom}=s'/s=M^2/s$ the fraction of 
the momentum of the proton carried
by the pomeron,
$M$ the diffractive mass and $s_0$ an energy scale not specified by the 
theory.

In analogy with Eq.~(\ref{total}), the term in brackets in 
(\ref{diffractive})
is identified as the $\pom-p$ total cross section, and therefore the factor 
\begin{equation}
f_{\pom /p}(\xi,t)\equiv \frac{{\beta_{\pom p}^2(t)}}{16\pi}\;\xi^{1-2\alpha(t)}
=\frac{{\beta_{\pom p}^2(0)}}{16\pi}\;\xi^{1-2\alpha(t)}F^2(t)
=K\;\xi^{1-2\alpha(t)}F^2(t)
\label{flux}
\end{equation}
is interpreted as a ``pomeron flux" (per proton) and used in calculating 
hard processes in diffraction dissociation in the 
Ingelman-Schlein model~\cite{IS}. 

Experimentally, 
the triple-pomeron coupling was found not to depend on $t$~\cite{KG} 
and therefore we will use $g(t)=g(0)$ and
$\sigma_T^{\pom p}(s',t)=\sigma_T^{\pom p}(s')$.

The function $F(t)$ represents the 
proton form factor. Donnachie and Landshoff proposed~\cite{DLF} 
that the appropriate 
form  factor for $pp$ elastic and diffractive scattering 
is the isoscalar form factor measured in electron-nucleon 
scattering, namely 
\begin{equation}
F_1(t)=\frac{4m^2-2.8t}{4m^2-t}\left[\frac{1}{1-t/0.7}\right]^2
\label{F1}
\end{equation}
where $m$ is the mass of the proton. When using 
this form factor, 
the pomeron flux 
is referred to as the Donnachie-Landshoff (DL) flux.\footnote{
The factor $K$ in the DL flux is  
$K_{DL}={(3\beta_{\pom q})^2}/{4\pi^2}$,
where $\beta_{\pom q}$ is the pomeron-quark coupling.}

In terms of the diffractive mass, the diffraction dissociation cross 
section is given by
\begin{equation}
\frac{d^2\sigma_{sd}}{dM^2dt}=
K\sigma_0^{\pom p}\;\frac{s^{2\epsilon}}
{(M^2)^{1+\epsilon}}\;\left(\frac{s}{M^2}\right)^{2\alpha' t}F^2(t)
\label{diffractiveM2}
\end{equation}
where $\sigma_0^{\pom p}=\beta_{\pom p}(0)\,g(0)/s_0^{\epsilon}$.  
The total diffractive cross section is given by
\begin{equation}
\sigma_{sd}(s)=K\sigma_0^{\pom p}\cdot s^{2\epsilon}
\displaystyle{\int_{M_0^2}^{0.1s}}\displaystyle{\int_0^{\infty}}
\frac{1}
{(M^2)^{1+\epsilon}}\;\left(\frac{s}{M^2}\right)^{2\alpha' t}F^2(t)
\,d\xi dt\;\sim s^{2\epsilon}
\label{totaldiffractive}
\end{equation}
where the lower limit of the $M^2$ integration 
is the effective diffractive threshold, $M_0^2=1.5$ GeV$^2$~\cite{KG},
and the upper limit, $M_{max}^2=0.1s$, corresponding to $\xi=0.1$,
is dictated by the coherence condition for diffraction~\cite{KG} 
(the contribution of the $M^2>0.1s$ region to the integrated cross section  
would be, in any case, negligibly small).
 
The $\sim s^{2\epsilon}$ dependence of $\sigma_{sd}(s)$ 
eventually leads to a diffractive cross section larger than the total, and 
therefore to violation of unitarity. 
The unitarity problem is a more general 
problem in pomeron pole dominance. It is well known that the  power law 
$s$-dependence of the total cross section, $\sim s^{\epsilon}$, violates the 
unitarity based Froissart bound, which  predicts that the 
total cross section cannot rise faster than  $\sim \ln^2 s$.  
Unitarity is also violated by the s-dependence of the ratio 
$\sigma_{el}/\sigma_T\sim s^{\epsilon}$, which eventually exceeds the 
black disc bound of one half ($\sigma_{el}\leq \frac{1}{2}\sigma_T$). 
However, in both the elastic and total cross sections 
 unitarization can be achieved by eikonalizing 
the elastic amplitude~\cite{CMG}. 
\section{The renormalized pomeron flux model}
Attempts to unitarize the diffractive amplitude 
by eikonalization~\cite{GLM} or by including cuts~\cite{K} have met 
with only moderate success.
Through such efforts it became clear that these 
``screening corrections" affect mainly the normalization of the 
diffractive amplitude, but leave the $M^2$ dependence almost 
unchanged. Such a trend is clearly present in the data, 
as demonstrated~\cite{CDF}
by comparing the CDF diffractive differential $\bar p p$ cross 
sections at $\sqrt{s}=$546 and 1800 GeV  with $pp$ 
cross sections at $\sqrt{s}=20$ 
GeV. Motivated by these
theoretical results  and the trend observed in the data, 
a phenomenological approach to unitarizing the diffractive amplitude
was proposed~\cite{R}
based on ``renormalizing" the pomeron flux 
according to 
\begin{equation}
f_N(\xi,t)=D\cdot f_{\pom/p}(\xi,t)
\label{fluxN}
\end{equation}
where the factor $D$ is determined by setting
\begin{equation}
D\displaystyle{\int_{\xi_{min}}^{0.1}}\displaystyle{\int_0^{\infty}}
f_{\pom/p}(\xi,t)d\xi dt=1
\label{D}
\end{equation}
if the value of the integrated standard flux
exceeds unity (the limits $\xi_{min}$ and $0.1$ 
of the $\xi$-integration are related to the limits 
$M_0^2$ and $0.1s$ of the $M^2$-integration in Eq.~\ref{totaldiffractive}
by $\xi=M^2/s$). Such a normalization, which 
corresponds to {\em at most} one pomeron per proton, leads to interpreting the 
pomeron flux as a probability density simply describing the $\xi$ and $t$ 
distributions of the exchanged pomeron in a diffractive process. 

The integrated standard flux is given by
\begin{equation}
N(\xi_{min})=\displaystyle{\int_{\xi_{min}}^{0.1}}
\displaystyle{\int_0^{\infty}}f_{\pom/p}(\xi,t)d\xi dt 
\sim (\xi_{min})^{-2\epsilon}\sim s^{2\epsilon}
\label{fluxI}
\end{equation}
and therefore $D\sim s^{-2\epsilon}$. Thus, flux renormalization 
approximately cancels the $s^{2\epsilon}$
dependence in Eq.~(\ref{diffractiveM2}) resulting in a slower rise 
of the diffractive relative to the total  cross section with energy and
preserving unitarity.  Figure~1, which is an updated version 
(including more data)
of a figure presented in Ref.~\cite{R}, shows 
the total diffractive cross section as a function of $\sqrt{s}$ along with 
the predictions obtained using the
standard (dashed curve) and renormalized (solid curve) 
pomeron flux (the dashed-dotted 
curve is the prediction of the Erhan-Schlein model, which is discussed below).
Renormalized flux predictions 
of {\em differential}
cross sections also show good 
agreement with data~\cite{KGSX,KGDIS}. Finally, using the Ingelman-Schlein 
model~\cite{IS}, the renormalized pomeron flux predicts correctly the 
measured rates of hard diffractive processes~\cite{R,KGSX,KGDIS}.

Predictions of hard diffraction rates using the standard/DL flux are 
unreliable due to a theoretical uncertainty inherent in the 
flux normalization. 
In Eq.~(\ref{total}) it is seen that $\beta_{\pom p}(0)$ can only be 
determined from the experimentally measured total cross section 
in terms of the energy scale $s_0$, which, as mentioned above, is not given by 
the theory ($s_0$ is usually 
set to 1 GeV$^2$, the hadron mass scale, but this is only {\em a convention}).
Thus, the normalization of the standard flux is  
unknown and therefore only predictions for {\em relative} hard diffraction 
rates are possible, as for example for the ratio of 
diffractive dijet production at two different energies.

The normalization uncertainty is resolved in the flux renormalization 
model.  The energy scale, $s_0$, can be determined~\cite{R} 
by setting the flux 
integral to unity at $\sqrt{s}\approx 20$ GeV, where the total diffractive 
cross section turns from  rising as  $\sim s^{2\epsilon}$ to assuming 
a rather flat 
$s$-dependence, presumably due to the saturation of the pomeron flux 
(see~Fig.~1). For  $\sqrt{s}>20$ GeV, 
where the flux integral is unity, the normalization is self-determined from
Eq.~(\ref{D}). Thus, predictions for hard diffraction using the 
renormalized pomeron flux can be made not only for relative but also for 
absolute rates. The recently reported CDF diffractive $W$ and dijet 
production rates~\cite{Moriond,MDIS} 
are in excellent agreement with the renormalized flux 
predictions~\cite{R,KGSX,KGDIS}.

\section{The Erhan-Schlein model}

Erhan and Schlein have taken a different approach to solving the 
unitarity problem of the triple-pomeron amplitude~\cite{PSSX,PSDIS}. 
They introduce a factor $D(\xi)$ in the flux to damp the small-$\xi$ values 
and thus slow down the $\sim s^{2\epsilon}$ dependence of the flux integral.
The normalization of the flux is left unchanged. 
To slow down the rise with $s$ of the {\em differential} cross section
at the higher $\xi$-values that are not affected by the damping factor,
they introduce a $\pom \pom R$ term, whose contribution increases at low 
energies due to the $\sim 1/\sqrt{s}$ dependence of the $R$-term. 
To fit ISR and UA8 data at $|t|\sim 1-2$ GeV$^2$, two more parameters are 
introduced. The detailed form of the proposed diffractive differential 
cross section is the following:
\begin{equation}
\frac{d^2\sigma_{sd}}{d\xi dt}
=f_{\pom p}(\xi,t)\cdot
\sigma_T^{\pom p}(s')
\label{ES1}
\end{equation}
\begin{equation}
f_{\pom p}(\xi,t)=D(\xi)\cdot K\cdot 
\xi^{1-2\alpha(t)}\cdot F_1^2(t)\cdot e^{bt}
\label{ES2}
\end{equation}
\begin{equation}
D(\xi)=\left\{\begin{array}{lc}
1&0.015<\xi<0.1~~\\
1-2700(\xi-0.015)^2&~10^{-4}<\xi<0.015\\
0.4-0.4\times 10^8(\xi-10^{-4})^2&~~~~~0<\xi<10^{-4}
\end{array}\right.
\label{ES3}
\end{equation}
\begin{equation}
\sigma_T^{\pom p}(s')=\sigma_0^{\pom p}\left[(\xi s)^{\Delta}
+r\cdot (\xi s)^{-0.45}
\right]
\label{ES4}
\end{equation}
\begin{equation}
\alpha(t)=1+\epsilon+0.26t+\alpha''t^2
\label{ES5}
\end{equation}
Using Eqs.~(\ref{ES1}-\ref{ES5}) and the {\em fixed} parameters
\begin{equation}
\epsilon=0.115 \;\;\;\;\;\Delta=0.08
\label{delta}
\end{equation}
diffractive cross section data are fitted with the following 
four free parameters determined from the data:
\begin{eqnarray}
C\equiv K\,\sigma_0^{\pom p}&=&0.73\pm 0.09\mbox{ mb GeV}^{-2}\\
b&=&0.75\pm0.27\mbox{ GeV}^{-2}\\
\alpha''&=&0.075\pm0.017\mbox{ GeV}^{-4}\\
r&=&5.0\pm0.6
\end{eqnarray}
It is claimed that good agreement is obtained with ``all 
available data"~\cite{PSDIS}. 
Below, we comment on the effect of the extra parameters 
introduced in the standard triple-pomeron amplitude, compare the predictions 
of the  model with experimental $pp$ and $\bar pp$ 
data and highlight some of its 
experimental and theoretical implications.
\subsection{The parameters of the Erhan-Schlein model}
In addition to the introduction of the $\pom\pom R$ term and the parameters 
associated with it, several additional  
parameters are introduced to the standard 
triple-regge amplitude:
\begin{itemize}
\item A different pomeron intercept is used in the 
$\sigma^{\pom p}(s')$ term than in the flux factor (see Eq.~\ref{delta}). 
\item Two damping factors, $D(\xi)$, are used for two different $\xi$-ranges. 
Figure~(2a) shows $D(\xi)$ as a function of $\xi$.  The roller-coaster shape 
of this distribution is mapped into the double resonance like distribution 
of the production cross section for a fixed  diffractive mass at $t=0$  
as a function of energy, as shown in Fig.~(2b).
\item The effect of the parameter $b$ of the $e^{bt}$ term in Eq.~(\ref{ES2})
is shown by the solid line in Fig.~(2c).
In the region $1<|t|<2$ GeV$^2$ of the UA8 data, which have been
presented in comparisons with the predictions of this 
model~\cite{PSSX,PSDIS}, the $b$-parameter accounts for a $t$-dependent 
reduction of the cross section by a factor of $2-5$.
\item The effect of the parameter $\alpha''$, which introduces 
a dependence of $1/\xi^{2\alpha'' t^2}$, is shown in Figs.~(2c,~2d).
Figure~(2c) displays this factor for $\xi=0.05$, the average UA8 $\xi$-value, 
 as a function of $t$. Within the UA8 $t$-range, it varies from
1.5 to 6. The $\xi$-dependence as a function of $t$ for this factor 
is shown for various 
$t$-values in Fig.~(2d). Again, within the UA8 range of $t$ and $\xi$, 
the effect is seen to be quite large. 
\end{itemize}
\subsection{Comparison with experimental $pp$ and $\bar pp$ data}
In Fig.~1, which shows total $pp$ and $\bar pp$ 
diffractive cross sections as a function 
of $\sqrt{s}$, the dashed line represents the standard flux prediction, the 
solid line the renormalized flux prediction and the dashed-dotted line the 
Erhan-Schlein prediction. The latter exibits two turn-overs as 
$\sqrt{s}$ increases, one at 
$\sqrt{s}\sim 10$ GeV and the other at $\sim 200$ GeV. The $\xi_{min}\sim 1/s$
values corresponding to these energies are at the interfaces 
of the damping factors that are used in the model. Clearly, 
a single damping factor would not provide a good fit to the data.

Figure~(3) shows differential cross sections $\xi d^2\sigma/d\xi dt$
at $t=-0.05$ GeV$^2$ 
as a function of $\xi$ for fixed target $pp$ data at $\sqrt{s}=14$ and 20 
GeV~\cite{C} and for the CDF $\bar pp$ 
data at 546 and 1800 GeV~\cite{CDF}. The fixed 
target data are for masses above the resonance region and the CDF 
data are for $\xi$-values large enough not to be affected by the 
experimental resolution of the $\xi$-measurement. 
The dashed curves are fits to the data using the form 
\begin{equation}
A\xi^{1-2\alpha_{\pom}(t)}\cdot (\xi s)^{\epsilon}
+B\xi^{1-2\alpha_{\pi}(t)}\cdot\sigma_T^{\pi p}(\xi s)
\label{pompi}
\end{equation}
with $\alpha_{\pom}(t)=1.104+0.25t$~\cite{CMG} 
and $\alpha_{\pi}(t)=0.9t$. The first term in Eq.~(\ref{pompi})
is the form of the triple pomeron amplitude and the second term the 
form for reggeized pion exchange. Fits with $\pom\pom\pom$ and $\pi$-exchange 
terms have been shown to represent well the $pp/\bar pp$ diffractive 
data~\cite{KG,KGSX,KGDIS}. Using the renormalized flux, such fits with
{\em only one free parameter}, namely the triple-pomeron coupling constant, 
yield differential cross sections that are in good agreement with the data 
not only in shape but also in normalization~\cite{R,KGSX,KGDIS}.

The solid curves in Fig.~(3) were 
calculated using the Erhan-Schlein model. Three features of these 
curves are immediately apparent:
\begin{itemize}
\item At low energies (Figs.~3a,~3b), 
the Erhan-Schlein cross sections are 
falling sharply with energy. This behaviour is due to the $\xi^{-0.45}$ 
dependence of the $\sigma^{\pom p}(s')$ term in the $\pom\pom R$ amplitude 
(second term in Eq.~\ref{ES4}), 
which dominates the low energy cross 
sections in this model due to the $s^{-0.45}$ factor, 
as compared to the $\xi^{\Delta}$ dependence of the 
$\pom\pom\pom$ term (first term in Eq.~\ref{ES4}). 
\item At the high energies (Figs.~3c,~3d), the 
curves initially rise as $\xi$ decreases and 
then bend over and fall as $\xi$ decreases further.  The bending point 
is at $\xi=0.015$, which is the $\xi$-value where the $D(\xi)$ 
damping takes effect. 
\item At $\xi=0.035$ the predictions are in 
relatively good agreement with the data. However, this agreement is not 
very meaningful as the parameters of the model were determined by fitting data 
of cross sections as a function of energy at this particular value of 
$\xi$~\cite{PSSX,PSDIS}. The argument offered for using only data at $\xi=0.035$
in the fits is that this $\xi$-value is low enough for  
the data to be background-free, but also high enough so that 
the $\xi$-distributions are not 
distorted by the experimental resolution in measuring $\xi$. As mentioned 
above, however, none of the data points in Fig.~(3) are affected by resolution 
and therefore the high energy data 
in the region of $\xi<0.035$ provide a good testing ground for the model. 
Figures~(3c,~3d)
show that the model is not very successful in this region.
\end{itemize}
\subsection{Experimental and theoretical implications}
In addition to the experimental issues already discussed 
in connection with  $pp/\bar pp$ 
cross sections, damping the pomeron flux at small-$\xi$ 
has serious implications for the HERA deep inelastic scattering (DIS)
and photoproduction results. 
These results are in the range $10^{-4}<\xi<10^{-2}$, 
within which the damping factor $D(\xi)=1-2700(\xi-0.015)^2$ is in effect.
The DIS H1~\cite{H1DIS} and ZEUS~\cite{ZEUS} results and the H1 
photoproduction results~\cite{H1PH} are compatible with a pomeron flux 
$\sim 1/\xi^{1+2\epsilon}$ with $\epsilon\approx 0.11$. From $\xi=10^{-4}$ to 
$\xi=10^{-2}$ the damping factor $D(\xi)$ increases by a factor of 2.2 while 
$\xi^{-2\epsilon}$ with $\epsilon$ from Eq.~(\ref{delta}) 
decreases by a factor of 2.8. Therefore, if $\xi$-damping were indeed in effect,
the value of $\epsilon$ expected at HERA would be $\epsilon\approx 0.02$.
Thus, the HERA data contradict the small-$\xi$ damping hypothesis.

On the theoretical side, there is no reason to expect that 
screening corrections  should damp the cross section 
preferentially at small diffractive masses. In fact, as already mentioned, 
eikonalization leaves the $M^2$ distribution largely unchanged~\cite{GLM}.  
As for the other parameters introduced in the model, there is no obvious 
reason why $\Delta$ should be different from $\epsilon$, or, if it were 
different, that it 
should have the value 0.08; neither is there a reason why the term 
$e^{bt}$ should be needed in diffraction, since a corresponding term 
$(e^{bt})^2=e^{1.5t}$ is not needed in the form factor for elastic scattering.
Finally, the introduction of the term $\alpha''t^2$ in the pomeron trajectory
makes both the diffractive and the elastic cross sections blow up at 
large values of $t$.
\section{Conclusions}
In an effort to solve the unitarity problem inherent to the triple-pomeron 
description of the diffractive cross section, Erhan and Schlein  
introduce a $\xi$-damping factor, $D(\xi)$, 
that decreases the pomeron flux for $\xi<0.015$. Since for $\xi>0.015$ 
the flux is left unchanged, the differential 
cross section $d^2\sigma/d\xi dt$ given by the $\pom\pom\pom$
amplitude still rises as $\sim s^{\epsilon}$ in this region, while 
experimentally it is found to decrease as $\sim s^{-\epsilon}$. To 
balance the $s$-dependence, a $\pom\pom R$ term that varies with $s$ as 
$\sim s^{-0.45}$ is introduced in the model. However, as this term 
has a $\xi$-dependence sharper than that of the $\pom\pom \pom$ term 
(a factor 
of $\xi^{\epsilon}$ is replaced by $\xi^{-0.45}$), it is now more 
difficult to fit the $\xi$-distributions of the data.
To obtain better fits, more parameters are introduced. In addition to choosing 
$\Delta\neq \epsilon$, four free parameters are used. These do not include 
the three parameters needed for the damping factors 
(two $\xi$-values and the value 
$D=0.4$ at $\xi=10^{-4}$), which were chosen 
to make the model agree with the $s$-dependence of the total diffractive cross 
section. Including these three parameters and the parameter $\Delta$ in 
the list of free parameters, 
a total of 8 free parameters are used, not taking into account the {\em shape}
of the damping factors,
which was also chosen to optimize the fits.
Furthermore, Erhan and Schlein point out that their model does not agree 
with  data at $|t|>\sim 0.5$ GeV$^2$, 
which includes the UA8 data, unless the $\xi$-damping factor is not used
in this region~\cite{PSDIS}.
However, no prescription is offered as to what happens at the 
interface between $|t|<0.5$, where damping is needed, and $|t|>0.5$, 
where it is not. Presumably more free parameters, which are currently hidden,
would be required to take this effect into account.
Despite the large number of explicit and hidden free parameters, 
the predictions of the model 
are contradicted by both the $pp/\bar pp$ and the HERA data. 
The model is also 
theoretically unsound as it predicts the vanishing of the $t=0$ cross section 
for small and increasingly larger diffractive masses as the energy increases.

In conclusion, the proposed model is ill-defined 
(low versus high $|t|$ behaviour), 
theoretically inconsistent and unphysical, contradicts $pp/\bar pp$ and HERA 
experimental data, and has no predictive power for hard diffraction rates 
within the framework of the Ingelman-Schlein model.

\section{Acknowledgements}
I would like to thank Philip Melese, Anwar Bhatti, 
Jose Montanha, Kerstin Borras and Suren Bagdasarov for many useful discussions.
I gratefully acknowledge the contribution of Jose Montanha
in the preparation of  the figures.

\newpage
\vglue 1in
\centerline{\hspace{-0.5in}\psfig{figure=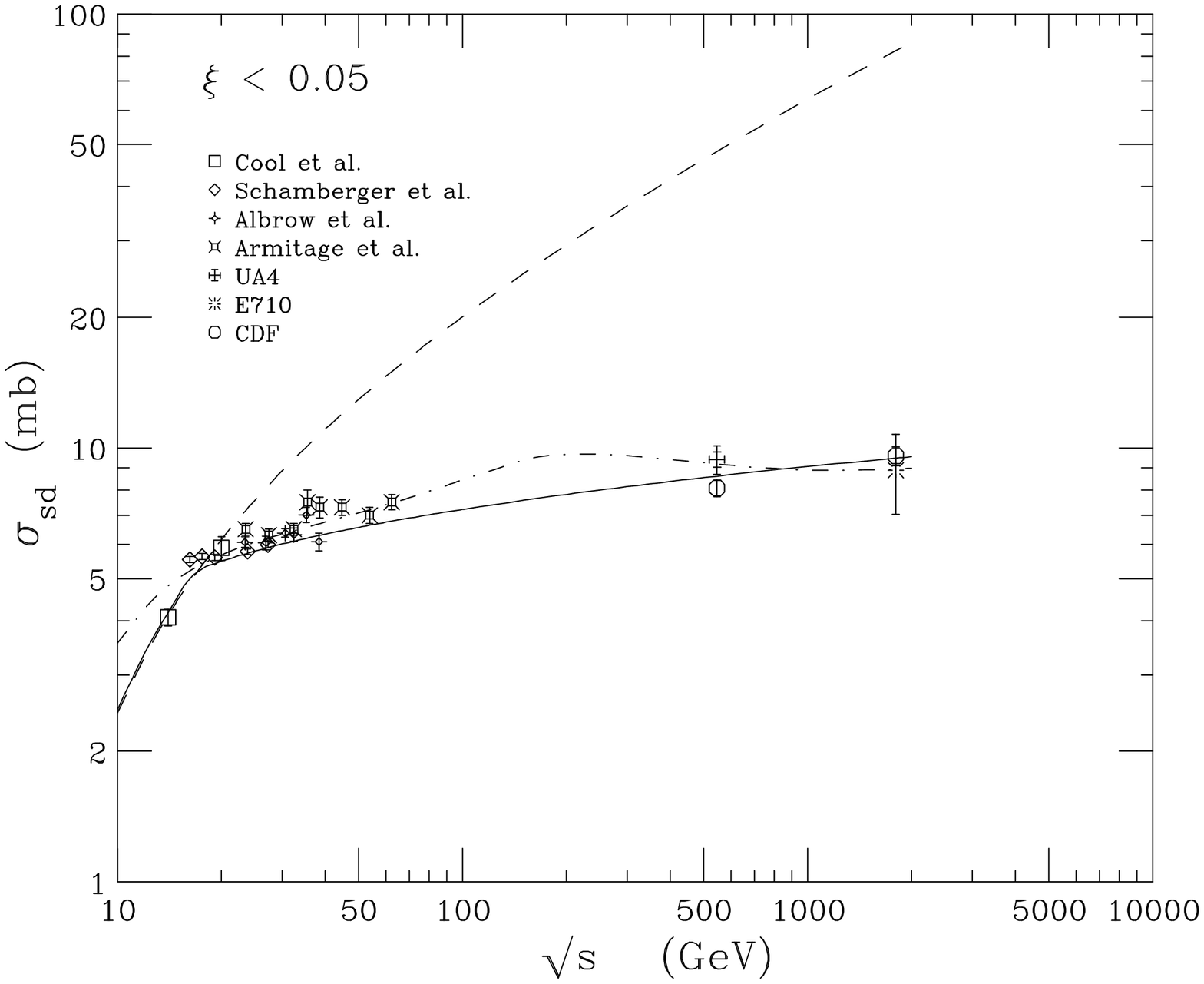,width=5in}}
\vfill
Figure 1:  The total $pp/\bar pp$ 
single diffraction dissociation cross section for $\xi<0.05$ as a function of  
center of mass energy. The dashed curve is the triple-pomeron prediction, the 
solid curve the renormalized flux prediction and the dashed-dotted curve the 
prediction of the Erhan-Schlein model of damping the pomeron flux 
at small-$\xi$. 
The latter exibits two ``turn-overs" as
$\sqrt{s}$ increases, one at
$\sqrt{s}\sim 10$ GeV and the other at $\sim 200$ GeV. The $\xi_{min}\sim 1/s$
values corresponding to these energies are at the interfaces
of the damping factors that are used in the model.
\clearpage
\newpage
\vglue 0.5in 
\centerline{\psfig{figure=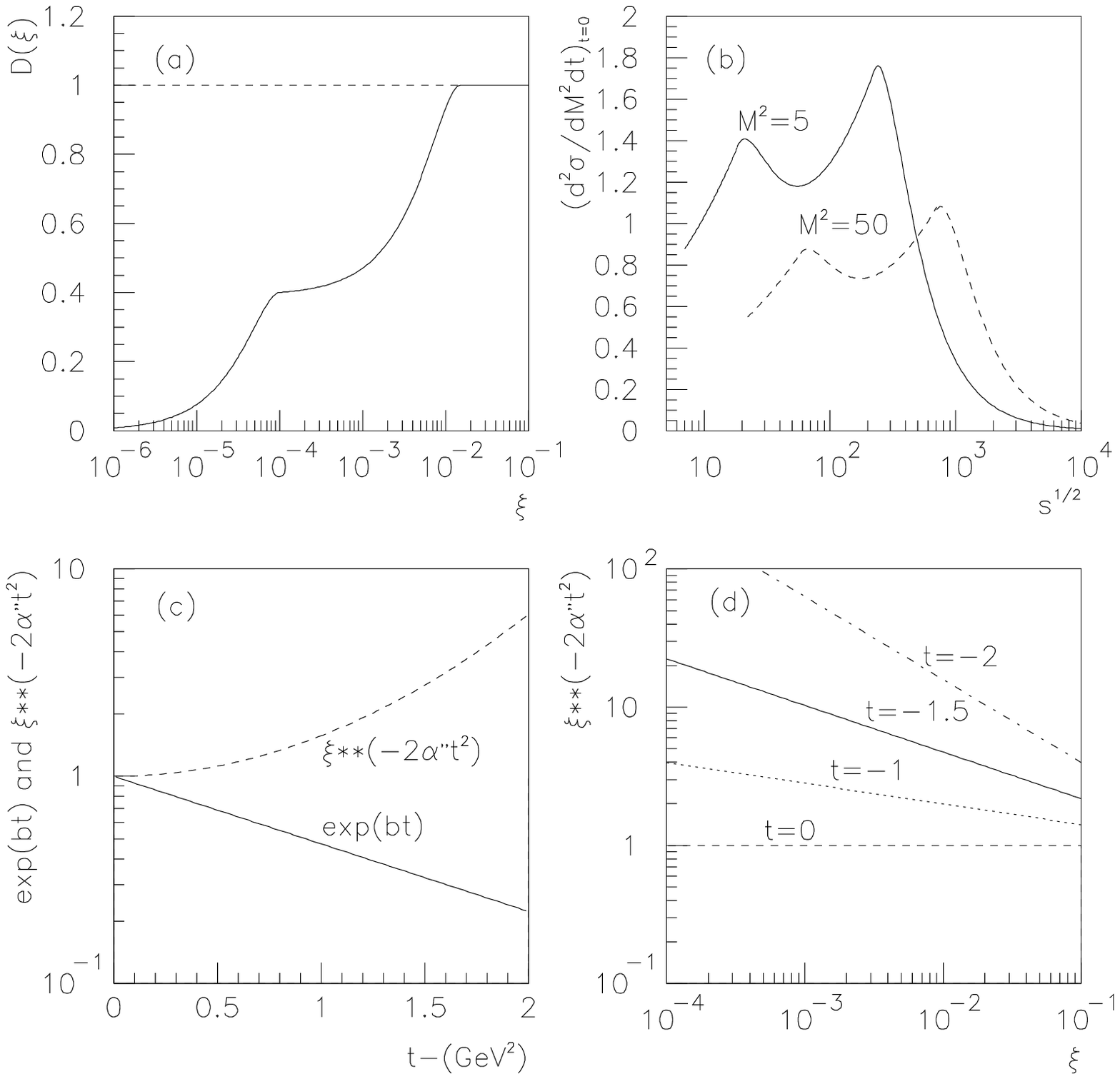,width=6.75in}}
\vfill
Figure 2:  The parameters of the Erhan-Schlein model: (a) the 
damping factor, $D(\xi)$, as a function of $\xi$; 
(b) $d^2\sigma/dM^2dt|_{t=0}$ versus $\sqrt{s}$ for $M^2=5$ GeV$^2$ (solid) 
and $10\times d^2\sigma/dM^2dt|_{t=0}$ for $M^2=50$ GeV$^2$ (dashed);
(c) the factors $e^{bt}$ (solid) and  
$\xi^{-2\alpha''t^2}$ at $\xi=0.05$ (dashed) 
versus $t$; (d) the factor $\xi^{-2\alpha''t^2}$ versus $\xi$ for 
$t=0,\, 1,\, 1.5$ and 2 GeV$^2$.
\clearpage
\newpage
\vglue 0.75in
\centerline{\psfig{figure=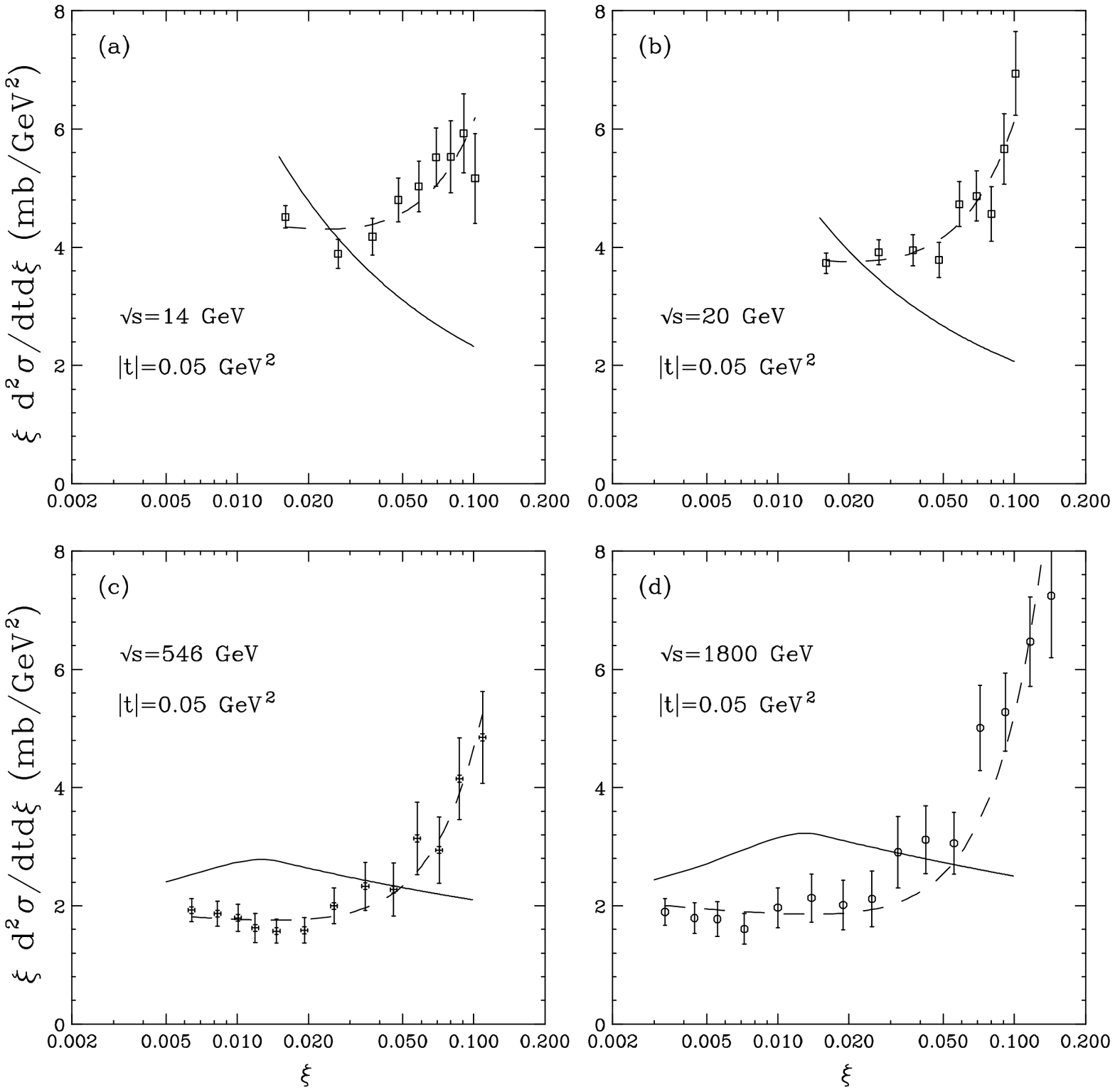,width=6in}}
\vfill
Figure 3: Cross sections $\xi d^2\sigma/dtd\xi$ versus $\xi$ for $pp/\bar pp$ 
data compared to theoretical predictions. The dashed lines are fits 
with a triple-pomeron term and a pion exchange term (see Eq.~\ref{pompi}). 
The solid lines represent the predictions of the Erhan-Schlein model. 
At the two highest energies, as $\xi$ decreases the solid curves  
are seen to ``bend-over" and fall 
at $\xi\approx 0.015$, where the $\xi$-damping factor becomes effective.
\clearpage
\end{document}